\documentclass[12pt]{article}
\usepackage{amssymb}
\usepackage{epsfig}

\catcode`\@=11
\def\numberbysection{\@addtoreset{equation}{section}
        \def\theequation{\thesection.\arabic{equation}}}

\def\beq{\begin{equation}}
\def\eeq{\end{equation}}
\numberbysection
\begin{document}
\begin{titlepage}
\begin{center}
\hfill  \\
\vskip 1.in {\Large \bf The scalar box integral and the Mellin - Barnes representation} \vskip 0.5in P. Valtancoli
\\[.2in]
{\em Dipartimento di Fisica, Polo Scientifico Universit\'a di Firenze \\
and INFN, Sezione di Firenze (Italy)\\
Via G. Sansone 1, 50019 Sesto Fiorentino, Italy}
\end{center}
\vskip .5in
\begin{abstract}
We solve exactly the scalar box integral using the Mellin-Barnes representation. Firstly
we recognize the hypergeometric functions resumming the series coming from the scalar integrals, then we perform an analytic continuation before applying the Laurent expansion in $\epsilon = (d-4)/2$ of the result. 
\end{abstract}
\medskip
\end{titlepage}
\pagenumbering{arabic}
\section{Introduction}

In 1974 't Hooft suggested that $QCD$ in the planar limit was exactly solvable \cite{1}. Unfortunately this program was too difficult to solve. There are however simpler quantum field theories in $d=4$ that could enjoy this property. In particular $N=4$ supersymmetric gauge theory is a cousin of $QCD$, ultraviolet finite, which is also a conformal field theory ( the one appearing in Maldacena's AdS/CFT correspondence \cite{2}). Can we solve $N=4$ super-Yang-Mills theory ? This is nowadays a hot topic, since its gravitational counterpart ( $N=8$ supergravity ) is also probably the only gravitational field theory which is ultraviolet finite.

There are good reasons to believe that, in the 't Hooft planar limit, higher loop orders of $N=4$ SYM are surprisingly simple. Indeed an iterative structure has been discovered at least in the four point planar amplitude, relating the two and three loop amplitudes to the one loop amplitude \cite{3}-\cite{4}-\cite{5} and eventually allowing the perturbative series to be resummed into a simpler result. This result has been found thanks to a work of Smirnov \cite{6}, which has evaluated the Laurent expansion in $\epsilon = (d-4)/2$ of the associated two-loop planar box integral. An important ingredient for his proof is the use of the Mellin-Barnes ( $MB$ ) representation of the double planar box integral. However it is somehow disappointing that this technique works only perturbatively in the infrared regulator $\epsilon$.

In this paper we analyze the $MB$ representation in all cases where a non-perturbative ( in $\epsilon$ ) amplitude can be computed, therefore concentrating our efforts to the four point one loop planar box integral, which is the basic information for the higher loops resummation. Also the arbitrary $N ( \ge 5 ) $point one loop amplitudes can be reduced to a sum over a set of basic scalar box four point integrals.

Back to the $MB$ representation, the non-perturbative point of view we follow allows us to rewrite the harmonic series coming from the one loop amplitude in terms of generalized hypergeometric functions ( of several variables ) which in the massless and massive cases ( up to $2$ external masses ) can be reduced to hypergeometric functions of a single variable.

A key ingredient in our method is that it is always necessary to perform an analytic
continuation of these hypergeometric functions before taking the Laurent expansion in $\epsilon$. We compare our findings with the direct method of integrating the scalar box and we find general agreement.

Our non-perturbative method could in principle be generalizable to the study of the hard two mass scalar box integral \cite{7} and eventually the double box scalar integral \cite{6}-\cite{8}, where the hard part is to find the correct analytic continuation of an hypergeometric function of three and four variables ( respectively ) which cannot be reduced to an hypergeometric function of a single variable. If we find the solution to this mathematical question we will report on it.

\section{Massless scalar box one loop integral}

Let us start with the massless scalar one loop box integral in $d$-dimensional space-time ( $\epsilon = ( d-2) / 4$:

\begin{eqnarray}  I & = & \Gamma( 2 - \epsilon ) \ \int^1_0 \ dx_1 \  dx_2 \ dx_3 \ dx_4 \ \delta( \ x_1 +  x_2  +  x_3  +  x_4  -  1 \ )  \nonumber \\
& & {[ \ x_1 x_3 ( -s ) \ + \ x_2 x_4 ( -t ) \ ]}^{\epsilon - 2}  \label{21} \end{eqnarray}

To proceed with the evaluation of the integral $I$ we choose the Feynman parameters as follows:

\begin{eqnarray}
x_1 & = & ( 1-x ) ( 1-y ) \nonumber \\
x_2 & = & x ( 1-y ) \nonumber \\
x_3 & = & y z \nonumber \\
x_4 & = & y ( 1-z )
\label{22}\end{eqnarray}

In this way the integral over $y$ factorizes and we end up with:

\begin{eqnarray}
I & = & \frac{ \Gamma^2(\epsilon) }{ \Gamma( 2 \epsilon )} \ \Gamma ( 2-\epsilon ) \ \int^1_0 \ dx \int^1_0 \ dz \ {[ \ ( 1-x) z (-s) \ + \ (1-z) x (-t) \ ]}^{\epsilon-2} \nonumber \\
&  & \int^1_0 \ dy \ {( y (1-y ))}^{\epsilon-1} = \frac{ \Gamma^2(\epsilon) }{ \Gamma( 2 \epsilon )}
\label{23}\end{eqnarray}

As we can see from eq. (\ref{23}), the integration over $x$ is straightforward and can be analitically performed. The resulting expression is:

\beq I \ = \ \frac{ \Gamma^2(\epsilon) }{ \Gamma( 2 \epsilon )}  \ \Gamma ( 1-\epsilon )
\ \int^1_0 \ dz \ \left[ \ \frac{ z^{\epsilon-1} ( -s )^{\epsilon-1} \ - \ (1-z)^{\epsilon-1} (-t)^{\epsilon-1} \ }{ ( (1-z) (-t) + z s ) } \right] \label{24}\eeq

By noting that the first integral over $z$ stands for the Euler representation of the hypergeometric function, we obtain the result:

\beq \int^1_0 dz \ z^{\epsilon-1} {( (-t ) - z ( - s - t ))}^{-1} ( -s )^{\epsilon-1} =
\frac{1}{\epsilon} \frac{  ( -s )^{\epsilon} }{  s \  t } {}_2 F_1 \left( 1 , \epsilon, \epsilon+1, 1 + \frac{s}{t} \right) \label{25}\eeq

The other integral can be easily evaluated by replacing $ z \rightarrow (1-z)$:

\beq \int^1_0 dz \ z^{\epsilon-1} {( (-s ) - z ( - s - t ))}^{-1} ( -t )^{\epsilon-1} =
\frac{1}{\epsilon} \frac{  ( -t )^{\epsilon} }{  s \  t } {}_2 F_1 \left( 1 , \epsilon, \epsilon+1, 1 + \frac{t}{s} \right) \label{26}\eeq

By summing eqs. (\ref{25}) and (\ref{26}) we finally find that the massless scalar box integral is:

\begin{eqnarray}
I & = & \frac{ \Gamma^2(\epsilon) }{ \Gamma( 2 \epsilon )} \frac{ \Gamma ( 1-\epsilon ) }{ \epsilon \ s \ t }  \  \left[  \ ( -s )^{\epsilon} \ {}_2 F_1 \left( 1 , \epsilon, \epsilon+1, 1 + \frac{s}{t} \right) + \nonumber  \right. \\
& + & \left.  ( -t )^{\epsilon} \  {}_2 F_1 \left( 1 , \epsilon, \epsilon+1, 1 + \frac{t}{s} \right) \right]
\label{27}\end{eqnarray}

The hypergeometric functions listed here admit the following Laurent expansion in the infrared regulator $\epsilon$:

\beq {}_2 F_1 \left( 1 , \epsilon, \epsilon+1, 1 + \frac{s}{t} \right) = 1 - \epsilon \log \left( -\frac{s}{t} \right) - \epsilon^2 Li_2 \left( 1 + \frac{s}{t} \right) + O ( \epsilon^3 ) \label{28}\eeq

producing the well known perturbative result

\beq I = \frac{2}{ s \ t \ \epsilon^2 } \frac{ \Gamma ( 1-\epsilon ) \Gamma^2 ( 1 + \epsilon ) }{ \Gamma ( 1 + 2 \epsilon ) } \ \left\{ \ (-s)^{\epsilon} \ + (-t)^{\epsilon}
\ + \ \epsilon^2 Li_2 \left( - \frac{s}{t} \right) \ + \
\epsilon^2 Li_2 \left( - \frac{t}{s} \right) - \epsilon^2 \frac{ \pi^2 }{ 3} \right\} \label{29}\eeq

that can be ultimately simplified by noting that

\beq  Li_2 \left( - \frac{s}{t} \right) \ + \ Li_2 \left( - \frac{t}{s} \right) \ = \
- \frac{1}{2} \log^2 \left( \frac{s}{t} \right) - \frac{\pi^2}{6} \label{210}\eeq

There is a second non-perturbative expression, owing to the following identity

\beq \frac{1}{\epsilon} \ {}_2 F_1 \ ( 1, \epsilon, \epsilon + 1, z ) \ = \
   \frac{1}{\epsilon} \ + \ \frac{1}{ 1 + \epsilon} \ z \
   {}_2 F_1 \ ( 1, 1 + \epsilon, 2 + \epsilon, z ) \label{211}\eeq

which allows to perform more easily the perturbative expansion in $\epsilon$

\begin{eqnarray}
I & = &  \frac{ \Gamma^2(\epsilon) }{ \Gamma( 2 \epsilon )} \
\left\{ \frac{\Gamma( 1-\epsilon )}{\epsilon  } \
\left( \ \frac{(-s)^{\epsilon}}{ s\ t} \ + \frac{(-t)^{\epsilon}}{s \ t}\ \right) + \right. \nonumber \\
 & + &   \frac{\Gamma( 1-\epsilon )}{ ( 1 + \epsilon ) } \
 \left[ \frac{(-s)^{\epsilon}}{ s \ t } \ \left( 1 + \frac{s}{t} \right) \ {}_2 F_1 \left( 1, 1+\epsilon, 2+\epsilon, 1 + \frac{s}{t} \ \right) \right. \nonumber \\
 &  + & \ \left.  \left. \frac{(-t)^{\epsilon}}{ s \ t }  \ \left( 1 + \frac{t}{s} \right)
 \ {}_2 F_1  \left( 1, 1+\epsilon, 2+\epsilon, 1 + \frac{t}{s} \right) \right]
 \right\} \label{212}\end{eqnarray}

 Also this second hypergeometric function can be developed in power series of $\epsilon$:

\beq \frac{1}{ 1 + \epsilon } \ z \ {}_2 F_1 ( 1, 1+\epsilon, 2+\epsilon,z) \
 = \ - \ \log ( 1-z ) \ - \ \epsilon Li_2 (z) \ + O(\epsilon^2) \label{213}\eeq

In this case we note that in the second part of eq. (\ref{212}) the term of order zero in $\epsilon$ cancels out while the term of order $\epsilon$ is given by

\begin{eqnarray} & - &  \ Li_2 \left( 1 + \frac{s}{t} \right) \ - \ Li_2 \left( 1 + \frac{t}{s} \right) \ - \
\log(-s) \log \left( -\frac{s}{t} \right) \ - \ \log(-t) \log \left( -\frac{t}{s} \right) \nonumber \\ & =  & Li_2 \left( - \frac{s}{t} \right) \ + \  Li_2 \left( - \frac{t}{s} \right) \ - \
    \frac{\pi^2}{3} \label{214}\end{eqnarray}

In this way the formula (\ref{29}) is recovered.

 \section{Mellin - Barnes representation of the massless scalar box integral}

We are going to compare the non-perturbative calculation of the massless scalar box (\ref{27}) with the $MB$ representation, a powerful method that has allowed to perform
analytically more complicated integrals, like the scalar double box integral.

The $MB$ representation allows us to rewrite the massless scalar box integral

 \beq  I \ = \ \frac{ \Gamma^2 ( \epsilon )}{\Gamma ( 2\epsilon)} \ \Gamma( 2-\epsilon ) \
 \int^1_0 \ dx \ \int^1_0 \ dz \ \left[ \ ( 1-x ) z \ (-s) + (1-z) x \ (-t) \
 \right]^{\epsilon-2} \label{31}\eeq

in the following way

 \begin{eqnarray} & & \Gamma ( 2-\epsilon )
  \ \left[ \ ( 1-x ) z \ (-s) + (1-z) x \ (-t) \ \right]^{\epsilon-2} = \nonumber \\
  & = & \frac{1}{2\pi i} \ \int^{+i \infty}_{-i\infty} \ dw \
  \Gamma( 2- \epsilon + w ) \ \Gamma(-w) \
  \frac{ ( \ ( 1-z ) \  x \ ( -t ) \ )^w  }{ ( \ (1-x) \ z \ (-s) \ )^{2-\epsilon+w} }
  \label{32}\end{eqnarray}

The integrals in $x$ and $z$ are now factorized and produce:

\begin{eqnarray}
\int^1_0 \ dx \ x^w \ (1-x)^{\epsilon-2-w} & = &
\frac{ \Gamma( w+1 ) \Gamma( \epsilon - 1 - w )}{ \Gamma ( \epsilon ) } \nonumber \\
  \int^1_0 \ dz \ (1-z)^w \ z^{\epsilon-2-w} & = &
  \frac{ \Gamma( w+1 ) \Gamma( \epsilon - 1 - w )}{ \Gamma ( \epsilon ) }
 \label{33}\end{eqnarray}

from which we obtain the $MB$ representation of the massless scalar box as

 \beq I = \frac{1}{\Gamma(2\epsilon)} \  \int^{+i \infty}_{-i\infty} \ \frac{dw}{2\pi i} \
 \frac{ (-t)^w }{ (-s)^{2-\epsilon+w} } \
 \Gamma^2 ( w+1 ) \ \Gamma( 2-\epsilon + w) \ \Gamma(-w) \ \Gamma^2 ( \epsilon -1 -w ) \label{34}\eeq

We can close the integration contour to the left or to the right and the result is unaffected by this choice. We always elaborate these integrals exactly without developing
in Laurent expansion of $\epsilon$; our point of view is therefore different from the usual applications of the $MB$ representation we have seen in literature.

For example, let us choose to analyze the poles to the left:

 1) it is more simple to analyze the simple poles $ w = \epsilon - 2 - n $; the corresponding series can be resummed into the following hypergeometric function

 \beq I_1 \ = \ \frac{\Gamma^2(\epsilon)}{\Gamma(2\epsilon)} \
 \frac{ \Gamma^2 ( 1-\epsilon )}{ \Gamma( 2-\epsilon )} \
 (-t)^{\epsilon-2} \ {}_2 F_1 \left( 1, 1, 2-\epsilon, - \frac{s}{t} \right) \label{35}\eeq

At fist sight this result is not much interesting, since this hypergeometric function doesn't admit a meaningful development in $\epsilon$ like those of the direct method.
It turns out that it is necessary to perform an analytic continuation into the right variables, before that developing in $\epsilon$ makes sense.

We are forced to choose the following analytic continuation

 \begin{eqnarray}
 {}_2 F_1 \ \left( 1, 1, 2- \epsilon, - \frac{s}{t} \right) & = &
 \left( -\frac{t}{s} \right) \frac{ \Gamma^2 ( \epsilon ) }{
 \Gamma ( 1 + \epsilon ) \Gamma ( \epsilon - 1) } \ {}_2 F_1 \
 \left( 1, \epsilon, \epsilon + 1, 1+ \frac{t}{s} \right) \nonumber \\
 & - &    \left( -\frac{t}{s} \right)^{1-\epsilon} \
 \frac{\Gamma^2 ( \epsilon ) \Gamma( 1- \epsilon)}{ \Gamma ( \epsilon - 1 )}
   \left( 1  +  \frac{s}{t} \right)^{-\epsilon}
\label{36}\end{eqnarray}

It then appears the first part of the standard solution obtained with the direct method, ( see eq. (\ref{27}))

   \beq \frac{ \Gamma^2 ( \epsilon ) }{ \Gamma ( 2\epsilon ) } \
   \frac{ \Gamma ( 1-\epsilon ) }{\epsilon } \ \frac{(-t)^{\epsilon}}{s \ t } \ {}_2 F_1 \
   \left(  1, \epsilon, \epsilon + 1, 1 + \frac{t}{s} \right) \label{37}\eeq

and a spurious term

   \beq - \ \frac{ \Gamma^3 ( \epsilon ) }{ \Gamma ( 2 \epsilon )}
   \frac{ \Gamma^2 ( 1-\epsilon ) }{ s \ t } \ s^{\epsilon} \
   \left( 1  +  \frac{s}{t} \right)^{-\epsilon}
   \label{38}\eeq

2) let us consider now the contribution of the double poles. It turns out that in order
to get a well defined resummed expression it is necessary to regularize the double poles
introducing a parameter $\delta$:

   \begin{eqnarray}
   I & = & \frac{1}{\Gamma ( 2\epsilon )} \ \lim_{\delta \rightarrow 0} \
   \int^{+ i \infty}_{-\infty} \ \frac{ dw }{ 2 \pi i} \frac{ (-t)^w }{ (-s)^{2-\epsilon+w }} \ \Gamma( w+1+\delta ) \ \Gamma( w+1 ) \ \Gamma( 2-\epsilon + w ) \ \Gamma(-w) \nonumber \\
   & & \Gamma( \epsilon - 1 - w ) \ \Gamma ( \epsilon - 1 - w - \delta ) \label{39}\end{eqnarray}

   Case 2a) first consider the simple poles $ w = - 1 - n - \delta $.
Their contribution can be easily resummed into the following formula

   \beq  I_{2a} \ = \ \frac{ \Gamma^2 ( \epsilon )}{ \Gamma ( 2 \epsilon )}
   \lim_{\delta \rightarrow 0} \ \Gamma(-\delta ) \ \Gamma ( 1+ \delta ) \
   \frac{ \Gamma ( \epsilon + \delta ) \Gamma ( 1 - \epsilon - \delta ) }{ \Gamma( \epsilon ) } \ \frac{ (-s)^{\epsilon} }{ s \ t } \
   \left( 1 + \frac{s}{t} \right)^{-\epsilon} \ \left( \frac{s}{t} \right)^\delta \label{310}\eeq

   A development in $ \delta $ is necessary

   \beq \Gamma ( \epsilon + \delta ) \ \Gamma ( 1 - \epsilon - \delta ) \ = \ \Gamma ( \epsilon ) \Gamma ( 1 - \epsilon ) \ ( 1 \ + \ \delta ( \psi ( \epsilon ) - \psi ( 1 - \epsilon ) ) ) + O ( \delta^2 ) \label{311}\eeq

   It follows that

 \begin{eqnarray}
 I_{2a} & = & - \  \lim_{\delta \rightarrow 0} \ \frac{1}{\delta} \ \frac{\Gamma^2(\epsilon)}{\Gamma ( 2\epsilon )} \ \Gamma ( 1-\epsilon ) \ \frac{ (-s)^\epsilon }{ s \ t } \ \left( 1 + \frac{s}{t} \right)^{-\epsilon} \ \nonumber \\
 & - & \frac{\Gamma^2(\epsilon)}{\Gamma ( 2\epsilon )} \ \Gamma ( 1-\epsilon ) \
 \left( \psi(\epsilon) - \psi(1-\epsilon) + \log \frac{s}{t} \right) \ \frac{
 (-s)^\epsilon }{ s \ t } \ \left( 1 + \frac{s}{t} \right)^{-\epsilon} + O ( \delta )
 \label{312}\end{eqnarray}

 $I_{2a}$ contains a singular term in $\delta$ and a finite term, which is the main contribution.

 Case 2b); we analyze now the simple poles $ w = - 1 - n $. Their contribution can be
 summarized as

 \beq I_{2b} \ = \ \frac{\Gamma^2 (\epsilon) }{ \Gamma( 2\epsilon )} \
 \lim_{\delta \rightarrow 0} \ \frac{1}{\delta} \ \frac{  \Gamma( \epsilon-\delta ) \Gamma ( 1-\epsilon ) }{ \Gamma ( 1-\delta ) \Gamma ( \epsilon ) } \ \frac{ (-s)^{\epsilon} }{ s \ t } \  {}_2 F_1 \ \left( 1, \epsilon - \delta, 1 - \delta, - \frac{s}{t} \right) \label{313}\eeq

Again we must choose a convenient analytic continuation in the variable $ ( 1 + \frac{s}{t} ) $:

\begin{eqnarray}
{}_2 F_1 \ \left( 1, \epsilon-\delta, 1-\delta, - \frac{s}{t} \right) & = &
\frac{ \Gamma ( 1-\delta ) \ \Gamma( -\epsilon ) }{ \Gamma ( -\delta ) \ \Gamma ( 1- \epsilon ) } \ {}_2 F_1 \ \left( 1, \epsilon - \delta, 1+ \epsilon
, 1+ \frac{s}{t} \right) \nonumber \\
& + & \frac{ \Gamma ( 1- \delta ) \Gamma ( \epsilon ) }{ \Gamma ( \epsilon - \delta ) }
\ \left( 1 + \frac{s}{t} \right)^{-\epsilon} \  {}_2 F_1 \left( -\delta, 1-\epsilon, 1-\epsilon, 1 + \frac{s}{t} \right) \nonumber \\
& &
\label{314}\end{eqnarray}

The firs term goes like $\delta$ and in this case it is possible to take directly the limit $\delta \rightarrow 0 $:

\beq I_{2b} \ \leftarrow \ \frac{ \Gamma^2 ( \epsilon ) }{ \Gamma ( 2 \epsilon )} \
\frac{ \Gamma ( 1-\epsilon ) }{ \epsilon } \ \frac{ (-s)^{\epsilon} }{ s \ t } \
{}_2 F_1 \ \left( 1, \epsilon , \epsilon + 1, 1 + \frac{s}{t} \right) \label{315}\eeq

representing the residual part of the exact solution ( eq. (\ref{27})).

The second term can be resummed as

\beq \lim_{\delta \rightarrow 0} \ \frac{1}{\delta} \  \frac{ \Gamma^2 ( \epsilon ) }{ \Gamma ( 2 \epsilon ) } \ \Gamma ( 1-\epsilon ) \ \frac{ (-s)^{\epsilon} }{ s \ t } \
\left( 1 + \frac{s}{t} \right)^{-\epsilon} \ \left( - \frac{s}{t} \right)^\delta \label{316}\eeq

that can be decomposed in a term divergent in $\delta$ plus a finite term

 \begin{eqnarray}
 & &  \lim_{\delta \rightarrow 0} \ \frac{1}{\delta} \frac{ \Gamma^2 ( \epsilon ) }{
 \Gamma ( 2 \epsilon ) } \ \Gamma ( 1-\epsilon ) \ \frac{ (-s)^{\epsilon} }{ s \ t } \
 \left( 1 + \frac{s}{t} \right)^{-\epsilon} \nonumber \\
 & + & \frac{ \Gamma^2 ( \epsilon ) }{
 \Gamma ( 2 \epsilon ) } \ \Gamma ( 1-\epsilon ) \ \log \left( - \frac{s}{t} \right) \ \frac{ (-s)^{\epsilon} }{ s \ t } \ \left( 1 + \frac{s}{t} \right)^{-\epsilon}
 \label{317} \end{eqnarray}

 Collecting our partial findings, we obtain with $MB$ representation the exact solution plus the following spurious terms

 \beq \frac{ \Gamma^2 ( \epsilon ) }{
 \Gamma ( 2 \epsilon ) } \ \Gamma ( 1-\epsilon ) \ \frac{ (-s)^{\epsilon} }{ s \ t } \
 \left( 1 + \frac{s}{t} \right)^{-\epsilon} \ \left\{ \ \log\left( - 1 \right) \ + \
 \Gamma(\epsilon) \Gamma( 1-\epsilon ) \ ( \ \cos (\pi \epsilon) \ - \ (-1)^{\epsilon} ) \right\} \ = \ 0 \label{318}\eeq

Surprisingly this lengthly expression is zero due to non trivial cancellation between all terms.

 \section{The scalar box integral with one external mass}

The next step is comparing the $MB$ representation with the direct method in the case of one external mass. To be precise we start with the following integral

  \begin{eqnarray}  I^m ( s, t, m)  & = & \Gamma( 2 - \epsilon ) \ \int^1_0 \ dx_1 \  dx_2 \ dx_3 \ dx_4 \ \delta( \ x_1 +  x_2  +  x_3  +  x_4  -  1 \ )  \nonumber \\
& & {[ \ x_1 x_3 ( -s ) \ + \ x_2 x_4 ( -t ) \ + x_1 x_4 ( - m^2 ) \ ]}^{\epsilon - 2}  \label{41}\end{eqnarray}

We choose the Feynman parameters as in the formula(\ref{22}) and we find the following expression

\beq
I^m \ = \ \frac{ \Gamma^2(\epsilon) }{ \Gamma( 2 \epsilon )} \ \Gamma ( 2-\epsilon ) \ \int^1_0 dx \int^1_0 dz \ {[ \ ( 1-x) \  z \ (-s) \ + \ (1-z) \ x \ (-t) \ + \ ( 1-x ) ( 1-z) ( - m^2 ) \ ]}^{\epsilon-2}
\label{42}\eeq

The integration in $x$ is again obvious and it gives rise to

\beq
I^m \ = \ \frac{ \Gamma^2(\epsilon) }{ \Gamma( 2 \epsilon )} \ \Gamma ( 1-\epsilon ) \ \int^1_0 dz \ \left[ \  \frac{ ( \ z ( -s ) \ + \ ( 1- z ) ( - m^2 ) \ )^{\epsilon-1}
- ( ( 1- z ) (-t) \ )^{\epsilon-1} }{ ( 1-z ) ( m^2-t ) \ + \ z s }
\right]
\label{43}\eeq

It is convenient introducing the following notations

 \beq z_0 = \frac{ m^2 - t }{ m^2 - t - s } \ \ \ \ z_1 = \frac{ m^2 }{ m^2 -s } \label{44}\eeq

 from which the integral $I^m$ we want to compute is of the form

 \begin{eqnarray}
 I^m & = & \frac{ \Gamma^2 ( \epsilon )}{\Gamma( 2\epsilon )} \ \Gamma( 1-\epsilon )
 \ \left\{ \
 \frac{ ( m^2 -s )^{\epsilon-1} }{ ( s + t- m^2 )} \ \int^1_0 dz \
 \frac{ ( z - z_1 )^{\epsilon-1} }{ z - z_0 }
 \right. \nonumber \\
 & - & \left. \
 \frac{ ( -t )^{\epsilon-1} }{ ( s + t- m^2 )} \ \int^1_0 dz \
 \frac{ ( 1 - z )^{\epsilon-1} }{ z - z_0 }
 \right\}
 \label{45}
 \end{eqnarray}

The first part can be rearranged as follows

 \begin{eqnarray}
 \int^1_0 dz \ \frac{ ( z - z_1 )^{\epsilon-1} }{ z - z_0 } & = &
 (-z_1)^{\epsilon-1} \
 \left( \frac{z_1}{z_1-z_0} \right) \ \int^1_0 dw \ w^{\epsilon-1}
 \left( 1 - \frac{ z_1 }{ z_1 - z_0 } w \right)^{-1} \nonumber \\
  & - &
 (1-z_1)^{\epsilon-1} \
 \left( \frac{1- z_1}{z_0-z_1} \right) \ \int^1_0 dw \ w^{\epsilon-1}
 \left( 1 - \frac{ 1-z_1 }{ z_0 - z_1 } w \right)^{-1} =
 \nonumber \\
 & = & \frac{(-z_1)^{\epsilon-1}}{\epsilon} \ \left( \frac{z_1}{z_1-z_0} \right)
 \ {}_2 F_1 \ \left( 1, \epsilon, \epsilon+1, \frac{z_1}{z_1-z_0} \right) \nonumber \\
 & - &  \frac{(1-z_1)^{\epsilon-1}}{\epsilon} \ \left( \frac{1- z_1}{z_0-z_1} \right)
 \ {}_2 F_1 \ \left( 1, \epsilon, \epsilon+1, \frac{1-z_1}{z_0-z_1} \right)
\label{46}\end{eqnarray}

By substituting eq. (\ref{46}) in $I^m$ and taking into account the definitions of
$z_0$ and $z_1$ in terms of $m^2, s, t$ we finally obtain

\begin{eqnarray}
I^m_1 & = & \frac{ \Gamma^2 ( \epsilon )}{\Gamma( 2\epsilon )} \ \frac{ \Gamma( 1-\epsilon) }{ \epsilon } \ \left\{
\ \frac{(-s)^\epsilon}{s \ t } \ {}_2 F_1 \ \left(
1, \epsilon, \epsilon+1, \frac{ s + t - m^2 }{ t }
\right) \right. \nonumber \\
& - & \left. \ \frac{(-m^2)^\epsilon}{s \ t } \ {}_2 F_1 \ \left(
1, \epsilon, \epsilon+1, \frac{ m^2 ( s + t - m^2)  }{ s \ t }
\right) \right\} \label{47}\end{eqnarray}

The second part of eq. (\ref{45}) is easily computed by replacing
$ z \ \rightarrow ( 1-z ) $

\beq I^m_2 \ = \  \frac{ \Gamma^2 ( \epsilon )}{\Gamma( 2\epsilon )} \ \frac{ \Gamma( 1-\epsilon) }{ \epsilon } \ \left[
\ \frac{(-t)^\epsilon}{s \ t } \ {}_2 F_1 \ \left(
1, \epsilon, \epsilon+1, \frac{ s + t - m^2 }{ s }
\right) \right]
\label{48}\eeq

The complete result is simply the sum of eqs.(\ref{47}) and (\ref{48})

\beq I^m \ = \ I^m_1 + I^m_2 \label{49}\eeq

In the limit $m^2 \rightarrow 0$ the massless scalar box integral $I(s,t)$ is found.

The result (\ref{49}) can be put in another form which makes more explicit the Laurent
expansion in $\epsilon$ of the non-perturbative solution

\begin{eqnarray}
I^m & = &  \frac{ \Gamma^2(\epsilon) }{ \Gamma( 2 \epsilon )} \
\left\{ \frac{\Gamma( 1-\epsilon )}{\epsilon  } \
\left( \ \frac{(-s)^{\epsilon}}{ s\ t} \ + \frac{(-t)^{\epsilon}}{s \ t}\  \ - \
\frac{ (-m^2)^{\epsilon}}{ s\ t} \ \right)
\right. \nonumber \\
 & + &   \frac{\Gamma( 1-\epsilon )}{ ( 1 + \epsilon ) } \
 \left[ \frac{(-s)^{\epsilon}}{ s \ t } \ \left( \frac{s + t - m^2}{t} \right) \ {}_2 F_1 \left( 1, 1+\epsilon, 2+\epsilon,  \ \frac{s + t - m^2}{t} \right) \right. \nonumber \\
 &  + & \ \frac{(-t)^{\epsilon}}{ s \ t }  \ \left( \frac{s + t - m^2}{s} \right)
 \ {}_2 F_1  \left( 1, 1+\epsilon, 2+\epsilon, \frac{s + t - m^2}{s} \right) \nonumber \\
 & - & \ \left.  \left. \frac{(-m^2)^{\epsilon}}{ s \ t }  \ \left( \frac{m^2 ( s + t - m^2 )}{s \ t} \right)
 \ {}_2 F_1  \left( 1, 1+\epsilon, 2+\epsilon, \frac{m^2 ( s + t - m^2 )}{s \ t} \right) \right]
 \right\} \label{410}\end{eqnarray}

In the expansion in $\epsilon$ it is enough to recall that

\beq \frac{1}{1+\epsilon} \ z \ {}_2 F_1 \ \left( 1, 1+ \epsilon, 2+\epsilon, z \right)
\ = \ - \log( 1-z ) \ - \ \epsilon \ Li_2 (z) \ + \ O ( \epsilon^2 ) \label{411}\eeq

The logarithmic terms cancels out between them, and at the order $\epsilon$ we find using the identity

\beq Li_2 ( 1-x ) \ = \ Li_2 (x) \ - \ \frac{\pi^2}{6} \ + \ \log(x) \ \log(1-x) \label{412}\eeq

the following contribution

\beq Li_2 \left( \frac{ m^2 - t }{s} \right) \ + \ Li_2 \left( \frac{ m^2 - s }{t} \right)
\ - \ Li_2 \left( \frac{ ( m^2 - s ) \ ( m^2 - t ) }{s \ t } \right) \ - \ \frac{\pi^2}{6}
\label{413}\eeq

\section{Comparison with the Mellin - Barnes integral representation}

The integral $I^m$ (eq.(\ref{41})) contains the sum of three terms and can be developed using the $MB$ representation by decomposing twice that sum

\begin{eqnarray}
& & \Gamma ( 2-\epsilon ) \ ( \ x_1 x_3 ( -s ) \ + \ x_2 x_4 (-t) \ + \ x_1 x_4 ( - m^2 ) \ )^{\epsilon-2} = \nonumber \\
& = & \int^{+ i \infty}_{- i \infty} \ \frac{ d\alpha}{2\pi i} \
\int^{+ i \infty}_{- i \infty} \ \frac{ d\beta}{2\pi i} \ \Gamma( -\alpha ) \ \Gamma( -\beta) \ \Gamma( 2 - \epsilon + \alpha + \beta ) \nonumber \\
 & & ( x_1 x_4 ( - m^2))^{\alpha} \
( x_1 x_3 ( - s))^{\beta} \ ( x_2 x_4 ( - t))^{\epsilon - 2 - \alpha -\beta}
\label{51} \end{eqnarray}

Introducing the usual Feynman parameters ( see eq. (\ref{22})) we find

\begin{eqnarray}
I^m & = & \frac{1}{\Gamma( 2\epsilon )} \ \int^{+ i \infty}_{- i \infty} \ \frac{ d\alpha}{2\pi i} \
\int^{+ i \infty}_{- i \infty} \ \frac{ d\beta}{2\pi i} \ \Gamma( -\alpha ) \ \Gamma( -\beta) \ \Gamma( 2 - \epsilon + \alpha + \beta ) \ \Gamma( \epsilon - 1 - \alpha - \beta ) \nonumber \\
& & \Gamma ( 1 + \beta ) \ \Gamma ( \epsilon - 1 - \beta ) \ \Gamma ( 1 + \alpha + \beta ) \ (-m^2)^{\alpha} \ (-s)^{\beta} \ (-t)^{\epsilon-2-\alpha-\beta}
\label{52}\end{eqnarray}

It is more convenient integrating firstly in $\alpha$ and then in $\beta$

\begin{eqnarray}
I^m & = & \frac{ (-t)^{\epsilon-2} }{ \Gamma ( 2\epsilon )} \
\int^{+ i \infty}_{- i \infty} \ \frac{ d\beta}{2\pi i} \ \left( \frac{s}{t} \right)^{\beta} \ \Gamma( - \beta ) \ \Gamma( 1 + \beta ) \ \Gamma( \epsilon - 1 - \beta ) \nonumber \\
& &  \int^{+ i \infty}_{- i \infty} \ \frac{ d\alpha}{2\pi i} \ \left( \frac{m^2}{t} \right)^{\alpha} \ \Gamma( -\alpha ) \ \Gamma( \epsilon - 1 - \alpha - \beta ) \
\Gamma( 2 - \epsilon + \alpha + \beta ) \ \Gamma( 1 + \alpha + \beta ) \nonumber \\
& & \label{53}
\end{eqnarray}

1) let us discuss the poles to the left of $\alpha$

\beq  \alpha = - 1 - k - \beta \label{54}\eeq

The corresponding series can be resummed as

\beq I^m_1 = \frac{ (-t)^{\epsilon} }{ t \ ( m^2 - t ) } \ \frac{ \Gamma( \epsilon ) \ \Gamma( 1 - \epsilon ) }{ \Gamma ( 2 \epsilon ) } \ \int^{+ i \infty}_{- i \infty} \ \frac{ d\beta}{2\pi i} \ \left( \frac{s}{m^2-t} \right)^{\beta} \ \Gamma ( - \beta ) \
\Gamma( \epsilon - 1 - \beta ) \ \Gamma^2 ( 1 + \beta ) \label{55}\eeq

Without loss of generality, we close the contour integral in $\beta$ to the right. There are two contributions

Case 1a: $ \beta = n $

\begin{eqnarray}
\tilde{I} & = & \int^{+ i \infty}_{- i \infty} \ \frac{ d\beta}{2\pi i} \
\left( \frac{s}{m^2-t} \right)^{\beta} \ \Gamma ( - \beta ) \
\Gamma( \epsilon - 1 - \beta ) \ \Gamma^2 ( 1 + \beta ) \rightarrow \nonumber \\
& \rightarrow & - \ \frac{ \Gamma ( \epsilon )}{ 1 - \epsilon } \ {}_2 F_1 \ \left(
1, 1, 2 - \epsilon, \frac{s}{m^2-t} \right)
\label{56}\end{eqnarray}

and

Case 1b: $ \beta = \epsilon - 1 + n $

\beq \tilde{I} \ \rightarrow \ \Gamma^2 ( \epsilon ) \ \Gamma ( 1 - \epsilon ) \
\left( \frac{m^2-t}{s} \right) \ \left( \frac{s}{m^2-s-t} \right)^{\epsilon}
\label{57}\eeq

By using the formula of analytic continuation

\begin{eqnarray}
{}_2 F_1 \ ( 1, 1, 2-\epsilon, z ) & = & \frac{\Gamma^2( \epsilon )}{\Gamma( 1+\epsilon )
\Gamma( \epsilon - 1 )} \ \frac{1}{z} \ {}_2 F_1 \ \left( 1, \epsilon, \epsilon + 1, 1 - \frac{1}{z} \right) \nonumber \\
& - & \frac{ \Gamma^2 ( \epsilon ) \Gamma ( 1-\epsilon )}{\Gamma ( \epsilon - 1 )} \
z^{\epsilon-1 } \ (1-z)^{-\epsilon}
\label{58}\end{eqnarray}

the whole contribution simplifies to

\beq \tilde{I} \ = \ \frac{\Gamma(\epsilon)}{\epsilon} \ \left( \frac{m^2-t}{s} \right)
{}_2 F_1 \ \left( 1, \epsilon, \epsilon+1, \frac{s+t-m^2}{s} \right) \label{59}\eeq

therefore the case $1$, which is the simplest one, results in

\beq I^m_1 \ = \ \frac{ (-t)^{\epsilon} }{ s \ t } \ \frac{\Gamma^2 ( \epsilon)}{\Gamma(2\epsilon)} \ \frac{ \Gamma( 1-\epsilon )}{\epsilon} \
{}_2 F_1 \ \left( 1, \epsilon, \epsilon+1, \frac{s+t-m^2}{s} \right) \label{510}\eeq

2) let us analyze the more difficult case, i.e. the series of poles

\beq \alpha = - 2 - k - \beta + \epsilon \label{511}\eeq

\begin{eqnarray}
I^m_2 & = & \frac{ ( -m^2 )^{\epsilon-2} }{ \Gamma( 2 \epsilon )} \ \sum_{k=0}^{\infty} \
(-1)^k \ \Gamma ( \epsilon - 1 -k ) \ \left( \frac{t}{m^2} \right)^k \nonumber \\
& & \int^{+ i \infty}_{- i \infty} \ \frac{ d\beta}{2\pi i} \
\left( \frac{s}{m^2} \right)^{\beta} \ \Gamma(-\beta) \ \Gamma( 1+\beta ) \ \Gamma( \epsilon-1-\beta ) \ \Gamma( k+2+\beta-\epsilon ) \label{512}\end{eqnarray}

Again we choose for simplicity to analyze the poles to the right

Case 2a: $\beta = n$

\beq I^m_{2a} \ = \ \frac{ ( -m^2 )^{\epsilon-2} }{ \Gamma( 2 \epsilon )} \ \Gamma^2 ( \epsilon -1 ) \Gamma( 2 - \epsilon ) \ \sum_{k,n=0}^{\infty} \
 \frac{ \Gamma( n+k+2-\epsilon ) \Gamma ( 2-\epsilon ) }{
 \Gamma( k+2-\epsilon ) \Gamma( n+2-\epsilon )} \ \left(\frac{t}{m^2}\right)^k
\ \left(\frac{s}{m^2}\right)^n \label{513}\eeq

This double series is tabulated \cite{9} and corresponds to a generalized hypergeometric function of two variables ( in particular of type $2$)

\beq I^m_{2a} \ = \ - \ (-m^2)^{\epsilon-2} \ \frac{ \Gamma(\epsilon) \Gamma(1-\epsilon)
\Gamma(\epsilon-1)}{\Gamma(2\epsilon)} \ F_2 \ \left( 2-\epsilon, 1, 1, 2-\epsilon, 2-\epsilon ; \frac{t}{m^2}, \frac{s}{m^2} \right) \label{514}\eeq

Let us note that this particular hypergeometric function of two variables can be reduced to a hypergeometric function of a single variable \cite{9}

\beq F_2 ( \alpha, \beta, \beta', \alpha, \alpha; x,y ) \ = \ (1-x)^{-\beta} \
(1-y)^{-\beta'} \ \ {}_2 F_1 \ \left( \beta, \beta', \alpha, \frac{ x \ y }{ (1-x) \ (1-y) } \right) \label{515}\eeq

from which we obtain

\beq I^m_{2a} \ = \ - \ \frac{ (-m^2)^{\epsilon} }{ ( m^2-t ) \ ( m^2-s ) } \
\frac{ \Gamma(\epsilon) \Gamma(1-\epsilon)
\Gamma(\epsilon-1)}{\Gamma(2\epsilon)} \ {}_2 F_1 \left( 1, 1, 2-\epsilon,
\frac{ s \ t }{ ( m^2-s) \ ( m^2-t )} \right) \label{516}\eeq

It is necessary applying again the analytic continuation ({\ref{58}), to obtain

\begin{eqnarray}
I^m_{2a} & = & - \ \frac{ (-m^2)^{\epsilon} }{ s \ t } \ \frac{\Gamma^2(\epsilon)}{\Gamma(2\epsilon)} \ \frac{\Gamma(1-\epsilon)}{\epsilon} \
{}_2 F_1 \ \left( 1, \epsilon, \epsilon+1, \frac{ m^2 ( s + t - m^2 )}{ s \ t } \right)
\nonumber \\
& + & \frac{ (-m^2)^{\epsilon} }{ s \ t } \ \frac{ \Gamma^3( \epsilon ) \Gamma^2(1-\epsilon)}{ \Gamma( 2\epsilon )} \ \left( \frac{ m^2 ( m^2 - s - t )}{ s \ t }
\right)^{-\epsilon}
\label{517}\end{eqnarray}

This formula contains a part of the exact solution plus a spurious term, as in the massless case.

Case 2b) let us analyze the poles $\beta= n + \epsilon -1$.

In this case the resummation is easier and results in

\beq I^m_{2b} \ = \ - \ \frac{ (-s)^{\epsilon} }{ s \ ( m^2 - s )} \
\frac{ \Gamma^2 ( \epsilon ) \Gamma ( 1-\epsilon )}{ \Gamma(2\epsilon) \ ( 1-\epsilon ) } \ {}_2 F_1 \ \left( 1, 1, 2-\epsilon, \frac{t}{m^2-s} \right) \label{518}\eeq

Applying again the analytic continuation formula ( eq.(\ref{58})), we obtain

\begin{eqnarray}
I^m_{2b} & = & \frac{ (-s)^{\epsilon} }{ s \ t} \
\frac{ \Gamma^2 ( \epsilon )}{ \Gamma(2\epsilon)}
\frac{ \Gamma ( 1-\epsilon )}{ \epsilon  } \
{}_2 F_1 \ \left( 1, \epsilon, \epsilon+1, \frac{ s+t-m^2 }{t}
\right) \nonumber \\
& - &  \frac{ (-s)^{\epsilon} }{ s \ t} \
\frac{\Gamma^3(\epsilon) \Gamma^2 ( 1-\epsilon ) }{\Gamma(2\epsilon)} \
\left( \frac{m^2 - s -t }{t}
\right)^{-\epsilon}
\label{519}\end{eqnarray}

To summarize, considering the sum of the three terms

\beq I^m \ = \ I^m_1 \ + \ I^m_{2a} \ + \ I^m_{2b} \label{520}\eeq

the spurious terms cancel out and we faithfully reproduce the exact result ( see eq. (\ref{49})).

To conclude the Mellin-Barnes method is completely equivalent to the direct method in the massive case, as in the massless case. We expect the same conclusion also in the case of easy two mass scalar box integral \cite{7}, whose exact result can be represented in terms of hypergeometric functions depending only on a single variable.

\section{Conclusions}

Using the Feynman parameter method, we have compared the Mellin-Barnes representation and the direct method of solving one loop scalar integrals at a non-perturbative level in the infrared regulator $\epsilon$. Usually the $MB$ representation is used in literature
perturbatively in $\epsilon$; in our approach we firstly recognize the hypergeometric functions resumming the series coming from the scalar integrals, then we perform an
analytic continuation in the right variables before applying the Laurent expansion in $\epsilon$ of the result.

Our method has been tested in the massless and massive ( with one external mass ) cases, but in principle it could be generalizable to more complicated integrals, like the hard two mass scalar box \cite{7} and the double scalar box \cite{6}-\cite{8}. Work is in progress in this direction.

\end{document}